\documentclass[conference]{IEEEtran}
\pdfoutput=1

\IEEEoverridecommandlockouts
\usepackage{cite}
\usepackage{amsmath,amssymb,amsfonts}
\usepackage{algorithmic}
\usepackage{graphicx}
\usepackage{textcomp}
\usepackage[table,xcdraw]{xcolor}
\usepackage{soul}
\usepackage{adjustbox}
\usepackage{multirow}
\usepackage{makecell}
\usepackage{hhline}

\def\BibTeX{{\rm B\kern-.05em{\sc i\kern-.025em b}\kern-.08em
    T\kern-.1667em\lower.7ex\hbox{E}\kern-.125emX}}
\begin{document}

\makeatletter
\newcommand{\linebreakand}{%
  \end{@IEEEauthorhalign}
  \hfill\mbox{}\par
  \mbox{}\hfill\begin{@IEEEauthorhalign}
}
\makeatother

\title{Health Guardian: Using Multi-modal Data to Understand Individual Health\\
}

\author
{\IEEEauthorblockN{
Vince S. Siu\IEEEauthorrefmark{1}\IEEEauthorrefmark{7},
Kuan Yu Hsieh\IEEEauthorrefmark{1}\IEEEauthorrefmark{5}\IEEEauthorrefmark{6}\IEEEauthorrefmark{7},
Italo Buleje\IEEEauthorrefmark{1}\IEEEauthorrefmark{7},
Takashi Itoh\IEEEauthorrefmark{2},
Tian Hao\IEEEauthorrefmark{1},\\
Ben Civjan\IEEEauthorrefmark{3},
Nigel Hinds\IEEEauthorrefmark{4},
Bing Dang\IEEEauthorrefmark{1}\IEEEauthorrefmark{8},
Jeffrey L. Rogers\IEEEauthorrefmark{1}\IEEEauthorrefmark{8},
Bo Wen\IEEEauthorrefmark{1}\IEEEauthorrefmark{8}}
\\
\IEEEauthorblockA{
\IEEEauthorrefmark{1}\textit{Digital Health, IBM T.J. Watson Research Center, Yorktown Heights, NY USA}\\
\IEEEauthorrefmark{2}\textit{Digital Health, IBM Research, Tokyo, Japan}\\
\IEEEauthorrefmark{3}\textit{Illinois Computer Science, University of Illinois Urbana-Champaign, Urbana-Champaign, IL, USA}\\
\IEEEauthorrefmark{4}\textit{Emerging Technology Engineering, 
IBM T.J. Watson Research Center, Yorktown Heights, USA}\\
\IEEEauthorrefmark{5}\textit{Department of Electrical and Computer Engineering, College of Electrical and Computer Engineering,}\\ \textit{National Yang Ming Chiao Tung University, Hsinchu 30010, Taiwan}\\
\IEEEauthorrefmark{6}\textit{Institute of Biomedical Engineering, College of Electrical and Computer Engineering,}\\ \textit{ National Yang Ming Chiao Tung University, Hsinchu 30010, Taiwan}\\
Emails: [\{vssiu, ibuleje, nhinds, dangbing, jeffrogers, bwen\}@us., kuan.yu@, jl03313@jp.]ibm.com\\
\{haotianrock, bencivjan\}@gmail.com\\
\IEEEauthorrefmark{7}Equal contribution\:
\IEEEauthorrefmark{8}Principal investigators\\
}}


\maketitle
\IEEEpeerreviewmaketitle

\IEEEpubidadjcol


\begin{abstract}
Artificial intelligence (AI) has shown great promise in revolutionizing the field of digital health by improving disease diagnosis, treatment, and prevention. This paper describes the Health Guardian platform, a non-commercial, scientific research-based platform developed by the IBM Digital Health team to rapidly translate AI research into cloud-based microservices. The platform can collect health-related data from various digital devices, including wearables and mobile applications. Its flexible architecture supports microservices that accept diverse data types such as text, audio, and video, expanding the range of digital health assessments and enabling holistic health evaluations by capturing voice, facial, and motion bio-signals. These microservices can be deployed to a clinical cohort specified through the Clinical Task Manager (CTM). The CTM then collects multi-modal, clinical data that can iteratively improve the accuracy of AI predictive models, discover new disease mechanisms, or identify novel biomarkers. This paper highlights three microservices with different input data types, including a text-based microservice for depression assessment, a video-based microservice for sit-to-stand mobility assessment, and a wearable-based microservice for functional mobility assessment. The CTM is also discussed as a tool to help design and set up clinical studies to unlock the full potential of the platform. Today, the Health Guardian platform is being leveraged in collaboration with research partners to optimize the development of AI models by utilizing a multitude of input sources. This approach streamlines research efforts, enhances efficiency, and facilitates the development and validation of digital health applications.

\end{abstract}


\begin{figure}[ht]
\centering
\includegraphics[width=\linewidth]{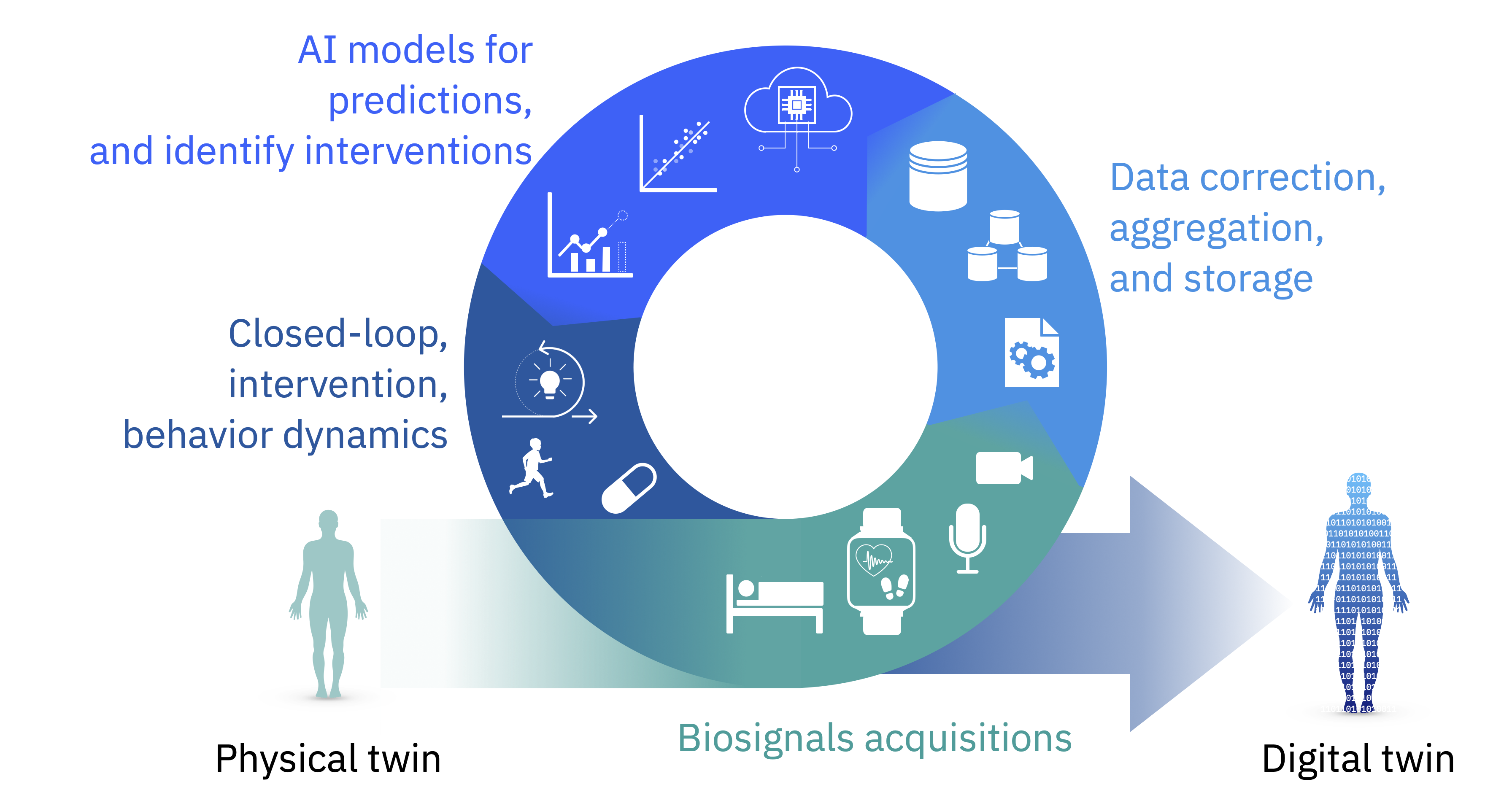}
\caption{Envisioned information flow diagram to create a virtual replica of a person, also known as a healthcare "digital twin”. Creation of a "digital twin" uses Internet-of-Things (IoT) technologies to acquire biosignals that are fed into AI-powered models to predict health insights for closed-loop interventions and/or behavioral changes.}
\label{Figure1}
\end{figure}

\begin{figure*}[ht]
\centering
\includegraphics[width=0.9\linewidth]{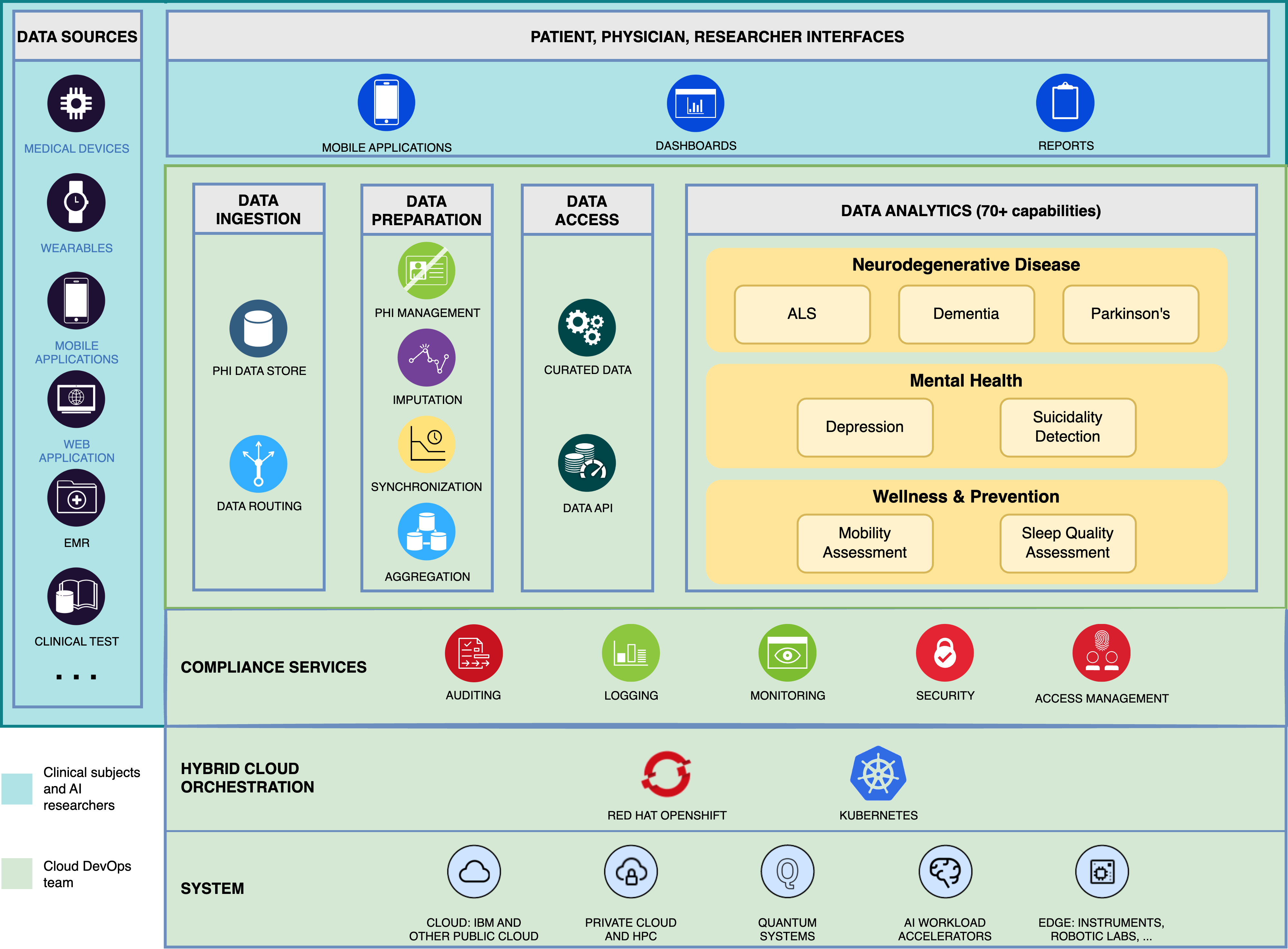}
\caption{Overview diagram of the Health Guardian Platform, highlighting the tools and services available in the data pipeline. This schematic is color-coded to denote the primary users: clinical subjects and AI researchers (cyan), and the Cloud DevOps team (green). In yellow are specific areas where AI/ML-based analytics have been developed and integrated into the Health Guardian Platform as a microservice. All components rest on top of a compliance services layer, and are built on an OpenShift and multi-cloud infrastructure.}
\label{Figure2}
\end{figure*}

\section{Introduction}
Digital Health is a growing interdisciplinary field that has seen a rise in popularity in recent years. The growth of personalized, predictive, and preventative healthcare has been fueled by the widespread use of mobile phones, Internet-of-Things (IoT) devices, and wearable sensors, as well as the affordability of cloud computing services. By incorporating new information technologies like edge computing, cloud computing, and artificial intelligence (AI) into healthcare, traditional practices can be improved, and innovative approaches can be created. For example, real-time monitoring of digital biomarkers using wearable and ambient technologies offers a comprehensive view of a person's health and are building blocks to creating a healthcare “digital twin”. The data from these digital biosignals can be collected and aggregated, then fed into AI and machine learning models. These models can identify important health insights and early interventions, which can then be shared with the patient and their healthcare team (Figure 1). AI research has already made significant progress in various healthcare domains, such as speech and language analysis for the diagnosis and monitoring of neurodegenerative diseases \cite{Norel2018,Yamada2020}, image analysis for automated detection of diabetic retinopathy \cite{Gulshan2016}, and assessments of gait \cite{Zou2020}, mobility \cite{Qi2018}, and drawing for evaluating cognitive decline in the elderly \cite{Kobayashi2022}. 

AI models that are trained with data from a multitude of input sources often yield better prediction results and 
performance. For example, evaluating cognitive decline by combining analysis of drawing and speech can provide a comprehensive understanding of both motor and linguistic aspects of cognition \cite{Yamada2021}. In diabetes management, sensors that can monitor the quality of sleep, activity levels, and appetite can improve prediction of daily insulin needs \cite{Karkkainen2022}. Fragmented remote patient monitoring systems make it difficult to gather data from various sensors with different input data types (e.g. audio, video, text, etc.). Integrating and coordinating data from these sensors and facilitating an easy way to collect patient data to train, validate, and oftentimes retrain, the models require significant coordination and time. To address these challenges, we introduce the Health Guardian (HG) platform, a comprehensive end-to-end solution that enables multi-modal assessments of an individual's health, and ensures secure data quality control at every stage of the data life cycle.

\section{Health Guardian Platform}

The Health Guardian (HG) platform provides a flexible framework for the rapid translation of AI research into microservices that can be used to collect and manage health-related data from clinical cohorts. Analytics developed using with AI and machine learning (ML) can be converted into deployable microservices using standard HG worker and API-gateway templates, as discussed in \cite{Wen2022}. The HG platform also allows users to create customized end-to-end data pipelines that test out the analytics, where data obtained from various microservices can be fed back into the AI predictive models for iterative improvements. A detailed description of the platform’s architecture and design is described in \cite{Wen2022}. 

\begin{table*}[ht]
\centering
\caption {Summary of microservices supported by Health Guardian}
\begin{adjustbox}{width=\textwidth,center=\textwidth} 
\renewcommand{\arraystretch}{2}
\begin{tabular}{w{c}{5em} c w{c}{8em} m{18em} w{c}{5em} w{c}{10em} c}
\hline
\multicolumn{1}{c}{Theme} &
  \multicolumn{1}{c}{\begin{tabular}[c]{@{}c@{}}Disease \\ Area\end{tabular}} &
  \multicolumn{1}{c}{Microservice} &
  \multicolumn{1}{c}{Description} &
  \multicolumn{1}{c}{\begin{tabular}[c]{@{}c@{}}Input File\\ (Format)\end{tabular}} &
  \multicolumn{1}{c}{Output} &
  \multicolumn{1}{c}{Ref} \\ 
  \hline
  &
  &
  \makecell{Timed Up \\and Go (TUG)} &
   Predict TUG score using data from daily walking activities that are passively captured by a smartwatch &
    \makecell{Text (.json)}  &
    TUG Score & 
    \cite{Hao2023}\\
    \hhline{~~-----}
  &
  &
  \cellcolor[gray]{0.9}\makecell{Sit-to-Stand} &
  \cellcolor[gray]{0.9}Analyze sit-to-stand either from a scripted or unscripted activity using an imager (Red-Green-Blue (RGB), depth, or millimeter wave camera) to extract metrics of mobility or motor symptoms and predict scores of standard mobility tests. &
  \cellcolor[gray]{0.9}\makecell{\\Video (.mp4)} &
  \cellcolor[gray]{0.9}\makecell{\\Torso Phase, \\ No. of Hesitations} & 
  \cellcolor[gray]{0.9}\cite{Mehta2021} \\
  \hhline{~~-----}
  &
  &
 \makecell{Bradykinesia} &
   Simple on-demand test to infer bradykinesia score from wrist worn gyroscope & 
  \makecell{Text (.json)}  &
  \makecell{Bradykinesia Score, \\ Pronation-Supination \\Score} & 
  \cite{Anand2020} \\  
  \hhline{~~-----}
  &
  \multirow{-12}{*}{\makecell{Parkinson's\\ Disease} }
   &
  \cellcolor[gray]{0.9}\makecell{Postural Instability \\and\\ Gait Disorder \\(PIGD)} &
  \cellcolor[gray]{0.9}Infers PIGD score using lumbar gyroscope/accelerometer from turns during a 1-min walk test. &
  \cellcolor[gray]{0.9}\makecell{Text (.csv)} &
  \cellcolor[gray]{0.9}\makecell{PIGD Score} & 
  \cellcolor[gray]{0.9}\cite{Anand2020} \\
  \hhline{~------}
 &
  \makecell{Amyotrophic\\Lateral\\Sclerosis} &
  \makecell{PsychE\\Acoustics} &
  Voice analysis to measure and predict progression of ALS. &
  \makecell{Audio (.wav)} &
  \makecell{ALS Score, \\Voice Report} &
  \cite{Norel2018}
  \\ \hhline{~------}
  &
   &
  \cellcolor[gray]{0.9}\makecell{PsychE\\Alzheimer’s} &
  \cellcolor[gray]{0.9}Voice analysis to measure and predict likelihood of developing Alzheimer’s Disease. &
  \cellcolor[gray]{0.9}\makecell{Audio (.wav)} &
  \cellcolor[gray]{0.9}\makecell{Likelihood Score of \\Developing Alzheimer’s} & 
  \cellcolor[gray]{0.9}\cite{Eyigoz2020}
  \\  \hhline{~~-----}
  \multirow{-24}{*}{\makecell{Neuro-\\ Degenerative\\Disease}} &
  \multirow{-2}{*}{\makecell{Alzheimer's\\ Disease}}&
  \makecell{Drawing} &
  Analyze freehand drawing with a digitizing tablet and pen to detect Alzheimer's disease and its prodromal stage (MCI) &
  \makecell{Image,\\Text (.zip)} &
  \makecell{Probability of cognitive \\ impairments, estimated \\ cognitive and clinical \\ measures including \\ neuro-pathological changes}
  & \cite{Kobayashi2022, Yamada2022}\\
  \hline
  \multirow{3}{*}{\makecell{Mental\\ Health}} &
  Depression &
  \cellcolor[gray]{.9}\makecell{PHQ-8 \\Depression\\ Questionnaire} &
  \cellcolor[gray]{.9}Provide standard PHQ-8 questionnaire through a mobile phone. &
  \cellcolor[gray]{.9}\makecell{Text (.json)} &
  \cellcolor[gray]{.9}PHQ-8 Score  & 
  \cellcolor[gray]{.9}\cite{KROENKE2009}
  \\ \hhline{~------}
  &
  \makecell{Suicidality} &
  \makecell{Suicidality\\Detection}  &
 Process the content of speech or input text and analyze to detect signs of depression and suicidality. &
  \makecell{Text (.json) \\Audio (.wav)} &
  Suicide Probability  &
  \cite{Agurto2018}
   \\  \hhline{-------}
\multirow{3}{*}{\makecell{Wellness \\ and \\ Prevention}} &
  Mobility &
  \cellcolor[gray]{.9}\makecell{Effective\\Mobility} &
  \cellcolor[gray]{.9}A mobility measure that accounts for different types of activity from walking to moving arms and hands. &
   \cellcolor[gray]{.9}\makecell{Text (.json)}& \cellcolor[gray]{.9}Effective Mobility Score
   &
   \cellcolor[gray]{.9}\cite{Bai2016}
    \\ \hhline{~------}
 &
  Sleep &
  \makecell{Sleep Quality}&
   Provide standard Stanford Sleep Questionnaire through a mobile phone.&
  \makecell{Text (.json)} &
  \makecell{Sleep\\Questionnaire\\Score}  &
  \cite{Shahid2011} 
   \\ \hhline{~------}
    &
  \makecell{Driving} &
  \cellcolor[gray]{.9}\makecell{Driving Risk\\Assessment}  &
 \cellcolor[gray]{.9}Extract speech features related to future risks of driving accidents by analyzing conversational speech with an AI chatbot. &
  \cellcolor[gray]{.9}\makecell{Text (.json) \\Audio (.wav)} &
  \cellcolor[gray]{.9}\makecell{Speech feature scores\\related to future\\driving accident risks}  &
  \cellcolor[gray]{.9}\cite{Yamada2021a}
   \\  
   \hline
\end{tabular}
\end{adjustbox}
\end{table*}

\begin{figure*}[ht]
\centering
\includegraphics[width=0.80\linewidth]{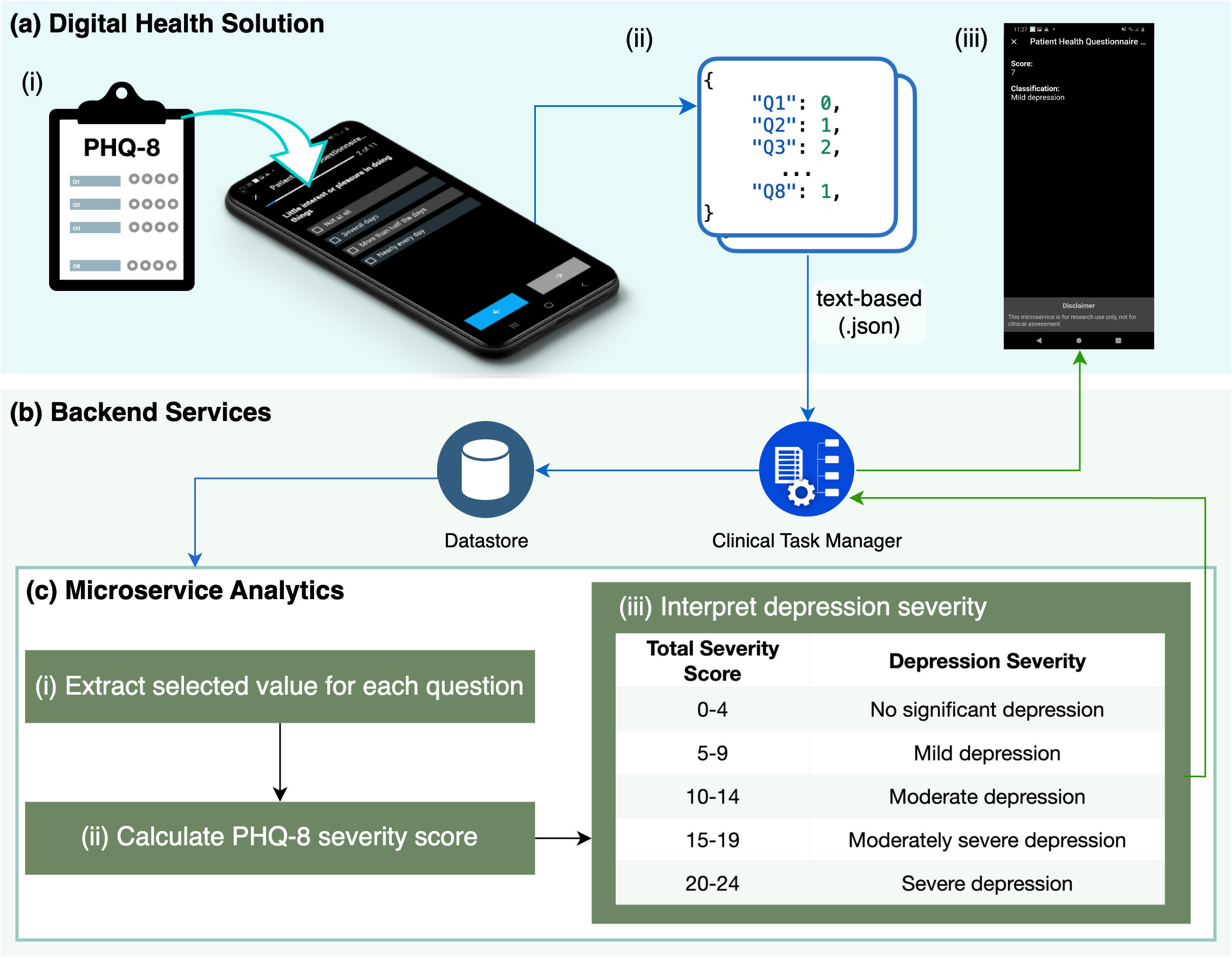}
\caption{(a) Illustrates the digital health solution for the PHQ-8 depression assessment microservice. Briefly, the user is prompted to provide responses to eight standard questions on the PHQ-8 questionnaire. (b) Depicts the Clinical Task Manager (CTM) and the Datastore which receives the input text-based responses in a (.json) file format after it has been uploaded from the mobile application. (c) Shows the scoring scale for the PHQ-8 questionnaire.}
\label{Figure3}
\end{figure*}

The data pipeline consists of five primary stages: data source, data ingestion, data preparation, data access and data analytics (Figure 2). First, data are collected from various data sources such as mobile and IoT devices, wearables, or electronic health records (EHRs). Then, the data are ingested into the clinical task manager (CTM) and routed into appropriate datastores. The data are then prepared using various strategies and steps, including de-identifying protected health information/personally identifiable information (PHI/PII), imputing missing data \cite{Schmitt2015}, synchronizing timestamps of data from different sensors, and aggregating data from multiple input streams. The pre-processed data are then accessed by the HG microservice workers to perform downstream analytics via data application programming interfaces (APIs). The HG platform currently supports over 70 capabilities in various focus areas, including neurodegenerative disease, mental health, and wellness and prevention. Finally, the results from the analytics can be viewed through various patient, clinician, or researcher interfaces, such as mobile applications, dashboards and reports.

The HG platform offers a versatile and reusable framework that can support AI/ML-based analytics with various input data types, including audio, text, video, and data from wearable or IoT devices, and can process data using either CPUs or GPUs. Table 1 showcases a subset of the microservices available on the HG platform, categorized by theme and disease areas. These microservices can be employed to manage and maintain individual health. For instance, in the aging population, where frailty, fall risks, and functional and cognitive decline are common, the HG platform can facilitate geriatric assessments and interventions both in-clinic and at-home. By selecting and deploying mobility and cognitive assessment microservices, patients can be evaluated comprehensively in different settings. Furthermore, the HG platform enables the study of Parkinson's disease patients over time through the deployment of microservices such as Timed Up and Go (TUG), sit-to-stand, bradykinesia, and the Postural Instability and Gait Disorder (PIGD). Additional microservices, such as the PHQ-8 depression questionnaire or suicidality prediction, utilizing audio or text-based inputs, can be added to provide a more comprehensive understanding of a subject's mobility and mental health status.

The following sections in this paper showcase three distinct microservices deployed on the HG platform. These microservices include a text-based microservice for depression assessment, a video-based microservice for sit-to-stand mobility assessment, and a wearable-based microservice for functional mobility assessment. For each microservice, we provide the clinical background, describe the digital health solution, and outline briefly the analytics used. Each one of these microservices can be deployed as a stand-alone assessment or combined with other microservices to obtain comprehensive insights into an individual's health. In the discussion section, we delve into the application of the CTM in creating clinical cohorts and facilitating the design and execution of clinical studies involving one or more deployed microservices. 

\section{Integration of Text-based Microservice for Depression Assessment}
\subsection{Clinical Background}
Depression is a prevalent mental health condition that affects many individuals worldwide. In 2020, an estimated 21.0 million (8.4{\%}) adults and 4.1 million (17.0{\%}) adolescents aged 12 to 17 in the United States alone, experienced at least one major depressive episode \cite{NIMH2020}. This condition can impact a person’s feelings, thoughts, and ability to perform daily activities, leading to persistent feelings of sadness, reduced interest in hobbies, and hopelessness lasting more than two weeks. In addition, depression can also cause physical symptoms such as changes in appetite and sleep patterns, fatigue and difficulty concentrating, thus affecting the individual’s capacity to work, sleep, study, and eat.

Validated self-report measures such as the Patient Health Questionnaire (PHQ-8) \cite{KROENKE2009}, the Beck Depression Inventory (BDI) \cite{beck1987} and the Depression Anxiety Stress Scales-21 (DASS-21) \cite{Henry2005} are commonly used for screening and assessing the severity of depression, guiding recommended treatments, and monitoring symptoms and recovery. However, traditional methods of administering these assessments are limited by the need for in-person, pencil-and-paper administration at a clinic, which can reduce the frequency at which individuals can be screened and monitored.

To address this issue, the IBM Digital Health team has developed a digital, text-based microservice using PHQ-8 depression questionnaire as a framework. The PHQ-8 consists of eight items that align with the diagnostic criteria for major depression disorder in the Diagnostic and Statistical Manual of Mental Disorders Fourth Edition (DSM-IV) \cite{KROENKE2009}, which is widely used for depression diagnosis and assessment. This digital microservice can increase access and the frequency at which individuals with mental health disorders can be screened and monitored. 

\begin{figure*}[ht]
\centering
\includegraphics[width=0.75\linewidth]{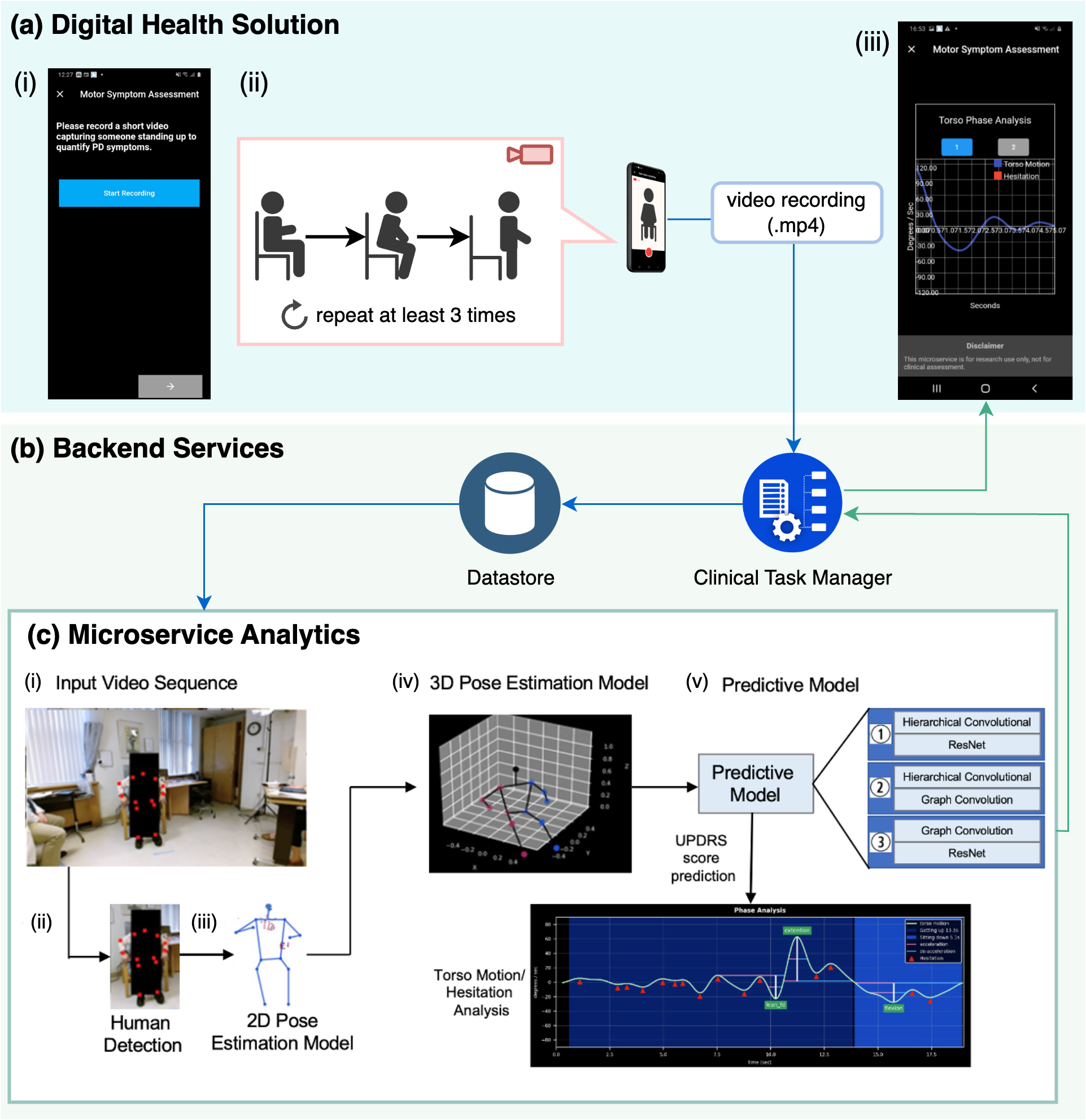}
\caption{(a) Illustrates the digital health solution for the sit-to-stand assessment microservice. Briefly, the user is prompted to record a short video of themselves cycling from a sitting to a standing position several times. (b) Depicts the Clinical Task Manager (CTM) and the datastore which are two backend services of the Health Guardian platform. These services receive the input video data containing a short sequence of sit-stand movements in .mp4 file format after it has been uploaded from the mobile application. (c) Describes the steps involved in the sit-to-stand microservice analytics that generate UPDRS score predictions and torso movement graphs from an input video.}
\label{Figure4}
\end{figure*}

\subsection{Digital Health Solution}
The PHQ-8 questionnaire is delivered to users via the Health Guardian mobile application (Figure 3a.i). Users can select the frequency of each depression symptom they have experienced in the past two weeks, and upon submission, the dataset is parsed into a text-based response in .json format (Figure 3a.ii), and uploaded to the clinical task manager (Figure 3b). 
The analytic worker processes the structured data, extracts the selected responses from each question (Figure 3c.i), and calculates the total score (Figure 3c.ii). The results are sent back to the mobile application via an API-gateway (Figure 3a.iii), where users can view and track their responses over time. The results are stored as structured data in a datastore, which can be connected to a clinician dashboard or other data rendering interfaces. These tools would enable clinicians or psychiatrists to observe changes in individuals' symptoms and treatment progression and make informed decisions regarding intervention strategies.

\subsection{Analytics and Validation}
The PHQ-8 questionnaire consists of eight items that evaluate symptoms of depression, with each item scored from 0 points (indicating "not at all" present) to 3 points (indicating "nearly every day" present). The total score can range from 0 to 24, with scores categorized as follows: no significant depressive symptoms (0 to 4), mild depressive symptoms (5 to 9), moderate (10 to 14), moderately severe (15 to 19), and severe (20 to 24) \cite{KROENKE2009}. 

Currently, the microservice provides a score for each completed PHQ-8 questionnaire (Figure 3a.iii), and stores the data longitudinally with time. Leveraging AI/ML models, it is possible to analyze the responses collected from multiple PHQ-8 questionnaires to derive more valuable insights. By considering longitudinal data, a deeper understanding of an individual's depressive symptoms and their progression can be obtained, enabling personalized and effective interventions.

Clinically, a variant of the PHQ-8, called the PHQ-9 questionnaire is used. The PHQ-8 differs from the PHQ-9 in that the PHQ-8 excludes the last question in the PHQ-9 which asks about thoughts of death and self-harm \cite{KROENKE2009}. The PHQ-8 has shown comparable ability to PHQ-9 in diagnosing and assessing depression among various populations \cite{RAZYKOV2012,Shin2019,WELLS2013}. Since the question about suicidal ideation is concerning when real-time psychiatric intervention or further suicidal evaluation are not provided \cite{Shin2019}, PHQ-8 may be a better alternative for depression screening when appropriate suicidal evaluation is not provided in the study or research.  

\section{Integration of Video-based Microservice for Sit-to-Stand Mobility Assessment}
\subsection{Clinical Background}

Advancements in computer vision and deep learning have enabled the development of video-based techniques for contact-free and passive evaluation of Parkinson's Disease (PD) symptoms during daily activities. PD is the second most common progressive neurodegenerative disease that affects 2-3{\%} of the population over 65 years of age \cite{Poewe2017}. Its hallmark characteristics are motor symptoms, such as bradykinesia (BRADY), tremors, rigidity, posture instability and gait disorders (PIGD), along with non-motor symptoms such as olfactory dysfunction and sleep disorders. Although no cure exists for PD, dopamine replacement therapy can mitigate symptoms and improve patients' quality of life.

Managing the various types of PD requires that neurologists comprehend the severity of symptoms and extent of motor fluctuations, which may require changes in medication timing and dosage. In-person assessments at a clinic done once or twice a year, follows protocols specified in the Unified Parkinson's Disease Rating Scale (UPDRS) and provide neurologists with information on the severity of symptoms, motor fluctuations, and medication efficacy.

In order to gain insights into changes in motor symptoms between clinical assessments, clinicians have asked PD patients to provide daily self-assessments or use wearable sensors to track mobility symptoms. However, recall bias limits the usefulness of self-reports, and compliance adherence issues can affect the deployment of wearable sensors. To address these challenges, IBM's Digital Health team has developed a video-based method that takes a short 20-30 second video containing sit-stand movements as input to predict the UPDRS subscores, such as BRADY and PIGD \cite{Mehta2021}.

\subsection{Digital Health Solution}
The sit-to-stand microservice is provided on the Health Guardian mobile application. Using the mobile application, the user is prompted to take a short video (\texttildelow20-30 s) of themselves cycling from a sitting to a standing position several times (Figures 4a.i and 4a.ii). The user uploads the video in .mp4 format from the mobile application to the clinical task manager (CTM) where the raw data and metadata of the file are stored in a database and cloud object storage (Figure 4b). The video file is processed by the analytic worker for this microservice to calculate the UPDRS scores, and generates a torso motion graph depicting the sit-stand movements and associated hesitations. The UPDRS scores and torso motion graph are sent back to the mobile application via an API-gateway (Figure 4a.iii).

\subsection{Analytics and Validation}
The analytics for the sit-to-stand microservice involves several steps and is described in detail in \cite{Mehta2021}. Briefly, a short input video sequence (Figure 4c.i) is processed by a human detector to extract the video frame-by-frame (Figure 4c.ii). The resulting data are then passed to a 2D pose estimation model that predicts the coordinate locations of human joints in 2D image space (Figure 4c.iii). Next, a 3D pose model utilizes the 2D pose information to predict joint locations in 3D Cartesian space (Figure 4c.iv). Finally, the 3D pose information is utilized in three different ensemble combinations that incorporate Hierarchical Convolutional Network (HCN) \cite{Li2018}, Spatio-Temporal Graph Convolutional Network (ST-GCN) \cite{Yan2018} and/or Convolutional Networks (CNNs) such as ResNet50 \cite{Caetano2019} (Figure 4c.v). The UPDRS score is predicted from the model, and graphs such as real-time torso motion can be generated from the processed data (Figure 4c.vi).

In two separate clinic visits, video clips of sit-stand motions from 35 subjects were captured and used to validate the analytics of the sit-to-stand microservice. This evaluation was part of a larger UPDRS assessment supervised by a neurologist, who was also assigned to score each task. The study demonstrated that it is possible to predict BRADY and PIGD scores from a short sit-stand video clip, with F1-scores, a measure of a model's accuracy, from the AI models performing better than the results from the two clinician video raters \cite{Mehta2021}. 

\section{Integration of Wearable-based Microservice for Functional Mobility Assessment}
\subsection{Clinical Background}

The Timed Up and Go Test (TUG) is a clinical assessment tool used to evaluate  balance and gait in everyday tasks such as sitting, standing, walking and turning. It is commonly used to examine functional mobility in older adults (aged 65+) who may be frail and have a history of falls \cite{Steffen2002}. The test involves standing up from a chair, walking 3 meters (10 feet), turning around, walking back to the chair, and sitting down again. The time taken to complete the test, measured in seconds, is strongly correlated to the level of functional mobility. 

Research has shown that older adults who took 13.5 seconds or longer to perform the TUG were at higher risks of falling, with a positive prediction rate of 90{\%} \cite{Shumway2000}. Additionally, studies have identified cutoff scores of 11.5 seconds in Parkinson's disease patients \cite{Nocera2013}, and 14 seconds for patients with strokes \cite{Andersson2006} as indicating increased fall risks.

The TUG test can also be used to monitor disease progression and changes in the quality of life for patients with mobility impairments as a result of specific diseases, as demonstrated in studies of patients with Parkinson's disease  \cite{Evans2017} and patients recovering from hip and knee arthroplasty \cite{Kennedy2006}. Although TUG is a valuable clinical tool, its limitations include the need for in-person assessment and the results are only measured as a snapshot in time.

\begin{figure*}[ht]
\centering
\includegraphics[width=0.75\linewidth]{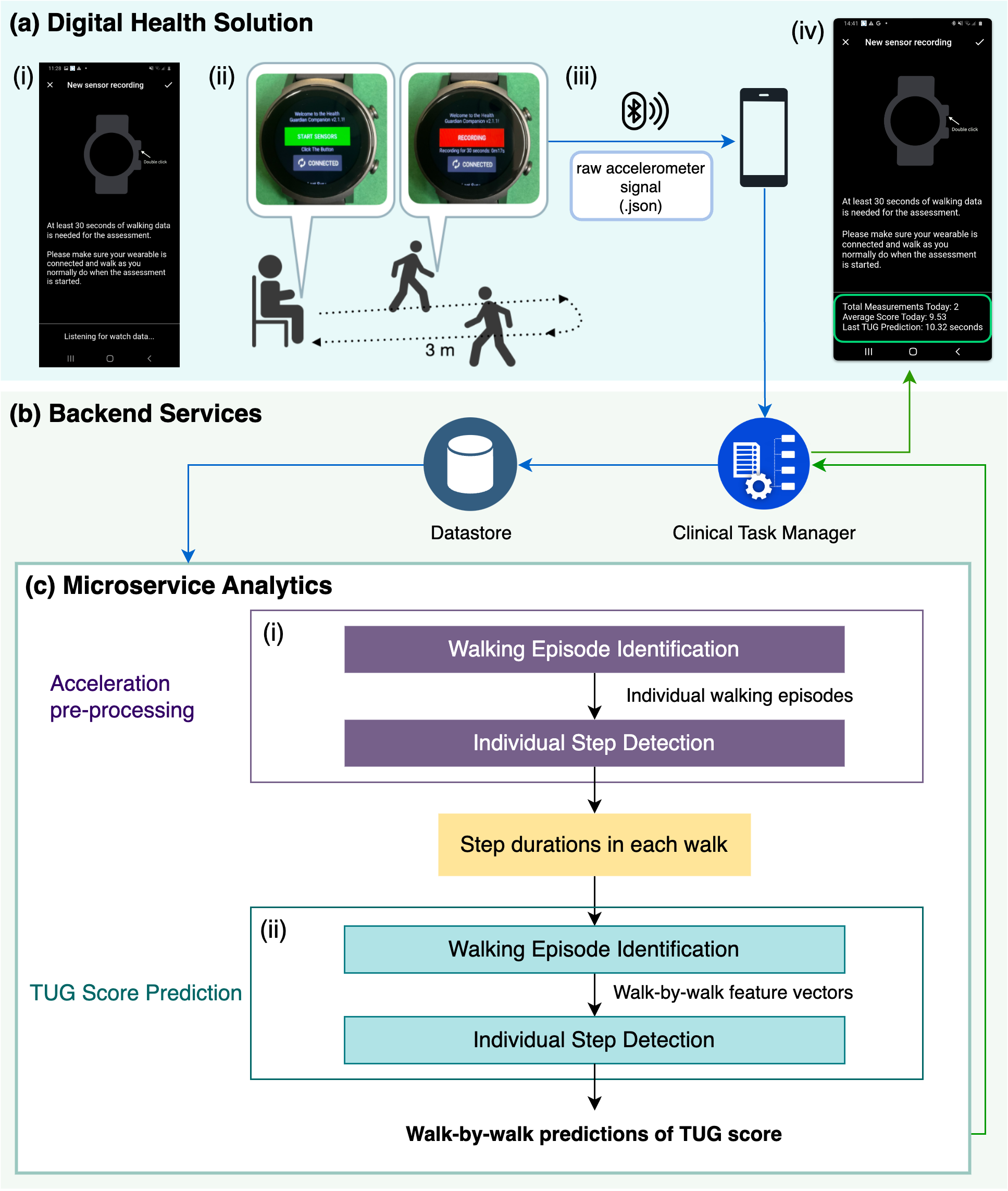}
\caption{(a) Illustrates the digital health solution for the Timed Up and Go (TUG) assessment microservice. Briefly, the user wears the Health Guardian companion watch and is prompted to press "start sensor" on the watch. The user will then stand up and walk for 30 seconds, then sit back down. The raw accelerometer signal from the watch is sent to the mobile application via Bluetooth. (b) Depicts the Clinical Task Manager (CTM) and the Datastore which receives the raw accelerometer signal in a .json file format after the file has been uploaded from the mobile application. (c) Describes the steps involved in the TUG wearable analytics and the TUG score prediction model.}
\label{Figure5}
\end{figure*}

\subsection{Digital Health Solution}

To adapt the TUG test to a digital health platform, IBM Research scientists have developed an automated TUG prediction AI model that utilizes accelerometer data from a wrist-worn watch. This model was transformed into a microservice using the standard HG worker and API-gateway template \cite{Wen2022}. 

To use this microservice, the subject first connects the HG mobile application to a wearable watch (e.g. TicWatch or Samsung Galaxy Watch series) that has the HG companion application installed. Next, the user selects the TUG microservice on the HG mobile application (Figure 5a.i), which prompts the user to press the "Start Sensor" button on the watch. The user then walks for at least 30 seconds until the watch buzzes to signal completion. The raw accelerometer signals are encrypted and sent in .json format via Bluetooth to the mobile application, which is then uploaded to the HG backend services for further processing (Figure 5b). The analytic worker processes the accelerometer data and calculates the TUG prediction score as well as the average TUG scores (if multiple walks were recorded in a given day) (Figure 5c). The results are then reported back to the mobile application via the microservice's API-gateway (Figure 5a.iv).

\subsection{Analytics and Validation}

The analysis framework for the TUG prediction model using wrist-worn accelerators is described in detail in \cite{Hao2023}. Briefly, the raw accelerometer data from the wrist-worn accelerator is pre-processed to identify walking episodes and to calculate step duration using step detection (Figure 5c.i). From the step durations and time difference between consecutive step durations, 20 statistical features are extracted for each identified walking episode using a Random Forest model (Figure 5c.ii). The predicted TUG score is derived from these statistical features, with the 25th and 5th percentile of step duration and the mean step duration having the greatest impact on the predicted TUG score. 

To validate this model, it was applied to three datasets that contained wearable recordings of walks and TUG scores from 303 subjects, including healthy individuals, those with Parkinson's disease, and those with mild cognitive impairment or dementia. The two public datasets used were the Long-term Movement Monitoring database (LTMM), which had subjects wear an accelerometer on their lower back, and the Gait in Parkinson's Disease (GPD) database, which recorded accelerometer data using in-sole gait sensors. The third dataset, the Dementia Behavioral Study dataset (DBSD), was collected by the University of Tsukuba and IBM Research and used wrist-worn accelerometers to record data during in-lab walking. The validation results demonstrated that the Random Forest-based predictive model for TUG had good clinical correlation, and achieved an accuracy of 1.7 +/- 1.7 seconds, with 84.8{\%} of the predictions falling within the minimal detectable change across all three separate cohorts \cite{Hao2023}.

\section{Discussion}
In the previous sections, three distinct microservices designed to assess depression, sit-to-stand mobility, and functional mobility using text, video, and wearable data, respectively were highlighted. While stand-alone microservices have clinical utility, recent research suggests that assessing clinical conditions with multitask conditions can lead to even better accuracy and assessments \cite{Shumway2000}. For instance, a study on older adults with balance issues found that while performing a secondary task, such as a language task, resulted in more swaying, suggesting that the effect of a secondary task on postural control depended on the subject's balance abilities, the difficulty of the balance task, and the type of secondary task being performed \cite{Shumway1997}.   

To support multi-modal microservice deployment on the Health Guardian platform in clinical studies, we have developed a Django-based web portal called the Clinical Task Manager (CTM) to manage study designs and patient cohorts. The web portal offers an easy-to-use graphical user interface for researchers and clinicians with limited programming background to leverage the full utility of the platform.

The CTM supports several key clinical study processes:
\begin{itemize}
\item Adding and managing subjects
\item Defining cohorts and assigning subjects
\item Defining tests and grouping them into test-sets
\item Dynamically defining tasks by assigning test-sets to cohort based on rules
\item Distributing tasks to edge devices for data collection
\item Providing multi-tenancy support by organizing subjects and tasks under separated study projects
\end{itemize}

The data model of the CTM is shown in Figure \ref{data-model}. To design a clinical study, researchers will first have to identify a patient population of interest, then use the CTM to filter and assign the subjects who meet the study enrollment criteria into a cohort (i.e. a subset of the subjects). Next, the researchers will need to identify one or more microservices that the subjects in the cohort will be asked to perform (i.e. PHQ-8, sit-to-stand, TUG, etc.). All the questions for each microservice is stored as a ‘test-set’ in the CTM. A study ‘task’ is formed when one or more test-sets is mapped to a cohort.  Once a task is created, the CTM will distribute the task to all edge devices of subjects in the assigned cohort with the HG mobile application installed. Subjects will be reminded to perform the tasks at the required time (e.g. within a certain time period after a subject wakes up if the task is to collect information about the subject’s sleep quality). 

\begin{figure}[htbp]
\centering
\includegraphics[width=0.85\linewidth]{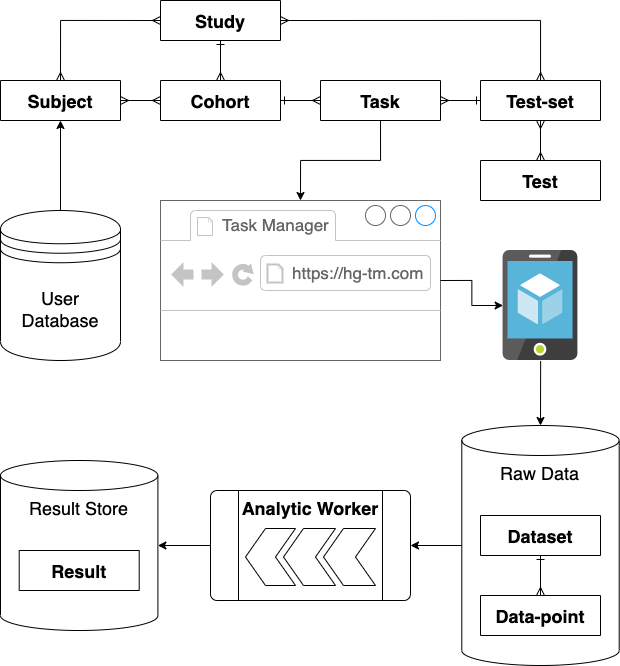}
\caption{Data model of the Clinical Task Manager (CTM). A 'study' includes a set of subjects, defined as a 'cohort', that are assigned a 'task' when a 'test-set' is mapped to the cohort. Subjects in the cohort will receive the 'task' on their Health Guardian mobile application. Upon successful completion of the task, the data are uploaded from each subject's phone to a backend datastore to be processed by the task's analytic worker. Results from the worker are stored in the datastore.}
\label{data-model}
\end{figure}

There are several unique features of this data model. One feature is that tasks can be assigned using rule-based dynamics. For example, if the goal of the clinical study is to explore the relationship between depression and mobility, the researcher can set up a rule where each day, a PHQ-8 task is sent to a cohort, and the responses can be filtered to create a sub-cohort with individuals whose score fall below a certain value. A subsequent task like sit-to-stand or TUG can be assigned and sent to this new sub-cohort.

Another feature is that this data model will generate a datapoint for each test completed by a subject. This datapoint can store a value, a string, or a data file along with the associated metadata like timestamps and user account information. These datapoints are grouped into datasets based on the test and test-set relationship. If one or more analytic pipelines are specified already, the system will automatically publish the datasets to the \textit{Orbit service}’s job queue for processing by the HG back-end components such as the analytic worker and API-gateways. A more detailed description of the \textit{Orbit service}, and setting up the analytic worker and API-gateways is provided in \cite{Wen2022}. After the analytics is completed, the analytic result will be stored into a database for researchers and clinicians to access and review.

The CTM is a critical component of the HG platform that facilitates an end-to-end solution for clinical study design in the digital health domain. By utilizing the CTM, researchers can effortlessly design clinical studies with one or more deployed microservices, and facilitate distributed data collection and processing. The CTM eliminates the common challenges typically associated with establishing and maintaining a data pipeline, enabling researchers to concentrate on data analysis and insight generation. New AI/ML models can be developed to conduct analytics and examine associations between health factors derived from data obtained and stored longitudinally from one or more microservices. The CTM streamlines the research process, empowering researchers to focus on extracting valuable insights from the data. 

\section{Conclusions}
In conclusion, we presented an overview of the Health Guardian platform, a comprehensive solution for collecting multi-modal data to gain insights into individual health. The HG platform simplifies cloud infrastructure and research components, providing standard worker and API-gateway templates to translate AI and ML-based predictive models into microservices. With 70+ capabilities, researchers and clinicians can design their clinical studies using various combinations of microservices such as the PHQ-8, sit-to-stand, and TUG described in this paper. The Clinical Task Manager's user-friendly graphical user interface allows for easy set up of clinical cohorts, and to select one or more microservices to assign to a specific cohort. The HG platform is flexible, scalable and supports the entire data life cycle, enabling accelerated development of AI research and clinical validation.

\section*{Acknowledgment}

The authors would like to acknowledge support from the IBM Research Accelerated Discovery Department.

\bibliography{Manuscript_2023_forArXiv}
\bibliographystyle{IEEEtran}

\end{document}